\pdfoutput=1 
\documentclass[12pt]{article}

\usepackage[margin=1in]{geometry}  
\usepackage{graphicx}              
\usepackage{amsmath}               
\usepackage{amsfonts}              
\usepackage{amsthm}                
\usepackage{url}                   

\DeclareMathSizes{12}{12}{11}{11}     

\newcommand{\sign}[1]{\textsf{#1}} 

\title{\textbf{Implementation of interaction between soft tissues and foreign bodies using modified voxel model}} 

\author{
\large
\textsc{Nikolaev S. N.}\\[2mm] 
\normalsize Saint-Petersburg State University \\ 
\normalsize mailto:  ser.niev@rambler.ru 
\vspace{-5mm}
}
\date{}

\begin{document}

article information: author -- Nikolaev S.N.; title -- Implementation of interaction between soft tissues and foreign bodies using modified voxel model;
journal title -- Computer tools in education journal; date -- December, 2013;
number -- 3; pages -- 28-32; url -- \url{http://ipo.spb.ru/journal/index.php?article/1581/};
journal publication language -- Russian.

\maketitle

\begin{abstract}
  Interactive bodies collision detection and elimination is one of the most popular task nowadays. Collisions can be detected in different ways. Collision search using space voxelization is one of the most fast. This paper describes improved voxel model that covers only area of collision interest and quickly eliminates collisions. This new method can be useful in real time collision processing of different rigid and soft bodies grids.

\end{abstract}

\section{Introduction}

\hspace{6mm}Modern medicine develops actively. Medical processes and materials become more perfect that helps surgeons to perform more qualitative and safer operations. Among all plastic and reconstructive operations with implants and expanders have a special importance. These operations allow achieving body symmetry and right proportions.

Nowadays the "virtual operation" principle has become a widely spread. In this principle an operated patient is modeled before operation that is need for planning. For example in case of implant or expander operation the implant with specific shape and sizes is selected by means of modeling result. This result also defines the implant placement method.

When implementing modeling operation the interaction handling problem between objects often appears. Different soft tissues are connected with each other so their interaction can be described via intermediate tissues properties. But some objects such as implants are alien; they are not connected with surrounding tissues. On the other hand models without interaction can simply penetrate each other. So in this case an approach for collision detection and handling is needed. The aim of this method is to detect models penetration area and to eliminate it. The collisions can be detected in several ways \cite{jimenez, lin, teschner}.

Collision handling methods can be split onto two groups \cite{faure}. Method based on restrictions calculates the common penetration area bound and moves models points abroad this area. At the same time method based on additional forces adds new forces (or extra speed which is almost the same) that try to revert models from collision conditions.

\section{Voxel model for collision handling}

\hspace{6mm}One of the most popular models for collision detecting is the voxel model which is more carefully described in \cite{gibson}. Initially voxels are used to describe object space position. Voxel model is based on creating grid connected with space (sometimes such grids creating are called Euler approach). The whole space is split into parallelepipeds that are called voxels. Each voxel can belong to some object in space. If it does not belong to any object this part of space is empty.

\renewcommand{\figurename}{Fig.}
\begin{figure}[ht!]
\centering
\includegraphics[width=90mm]{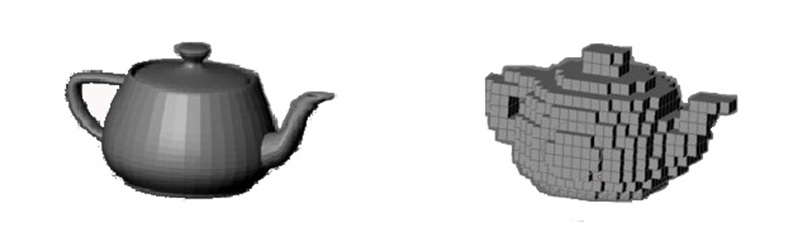}
\caption{\sign{Voxel model}}
\label{overflow}
\end{figure}

Any object in question fills in some part of space. If we take all voxels that belong to this object we get the whole object location in space exactly up to the size of voxel (Pic. 1).

\renewcommand{\figurename}{Fig.}
\begin{figure}[ht!]
\centering
\includegraphics[width=90mm]{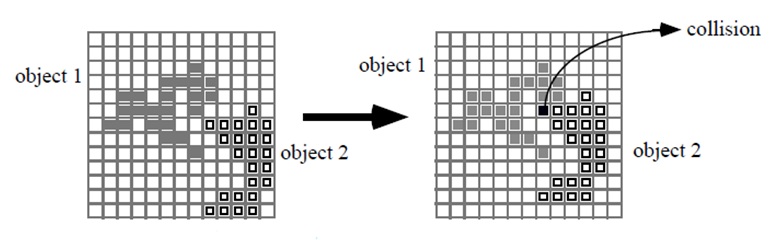}
\caption{\sign{Collision detection with voxels}}
\label{overflow}
\end{figure}

If any voxel belongs to two objects at once then these objects are in collision (Pic. 2). Also voxel model exactly shows where in space the collision takes place \cite{gibson}.

Compared with other approaches voxel model allows calculating collisions for bodies with random shape. This is important for soft bodies because collision detection does not depend on their changeable shape and sizes. Calculation simplicity is also voxels distinctive feature, thus collisions can be detected in real time.

Several papers describe voxel model implementation for soft bodies.

In works \cite{heide, schiemann} voxels are used together with surface triangle model. The triangle model is responsible for deformations while voxel model is responsible for collisions processing. The authors of paper give transformations between models and simulate system behavior using these transformations on each modeling iteration step.

Developers from \cite{faure} use voxel model to regard bodies penetration depth. Their approach works with every physical grid. Voxel model is built for space with physical bodies. The amount of joint voxels defines penetration depth. To eliminate collisions authors use method based on restrictions which paper also briefly describes.

\section{Voxel model modification}

\hspace{6mm}To handle collisions between soft tissues and alien bodies we have modified voxel model. Our modification is based on simple assumptions:

1. Only one voxel collision model is needed for two interactive models.

2. In interaction just bodies surfaces are in contact. So we can build voxel model only for area of interest. Also there is no need in penetration depth estimation.

3. Any soft body model is look like grid that connected with some kind of links. In this approach link types does not matter.

\begin{figure}[ht!]
\centering
\includegraphics[width=70mm]{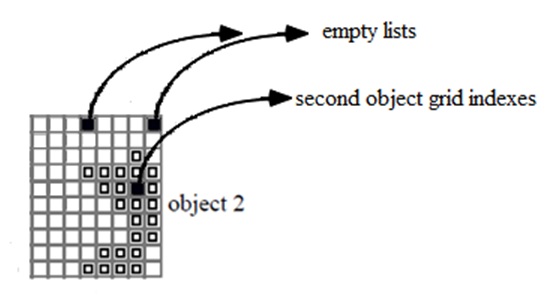}
\caption{\sign{Voxel model modification}}
\label{overflow}
\end{figure}

As a result we add a list in each voxel that store grid indexes for one of interacted models (Pic. 3). Each iteration step voxels lists are updated before collision search and elimination. The formula that define voxel index for given vertex in OX direction is look like this:
\begin{equation}
i_{\text{x}} = (x_{\text{v}} - x_{\text{0}}) / v_{\text{x}}
\end{equation}
where $i_{\text{x}}$ -- is voxel index, $x_{\text{v}}$ -- x element of vertex coordinate, $x_{\text{0}}$ -- x element of object bounding cube center coordinate, $v_{\text{x}}$ -- voxel size in OX direction. If calculated index value is more than voxels maximal indexes, we treat this index is beyond the bound and reject it. For speed improvement we take voxel sizes as multiples of two. The calculations for OY and OZ directions are performed similarly. Thus we define set of voxels that belongs to first contacting object (Pic. 3).

To find the set of belonging voxels for the rest object we execute the same calculations of voxel indexes. But in this case the indexes are not stored in voxel lists. Any index that refers to voxel with nonempty list detects a collision.

\section{Collision handling system architecture}

\hspace{6mm}Collision handling system architecture is presented on Pic. 4.

\begin{figure}[ht!]
\centering
\includegraphics[width=100mm]{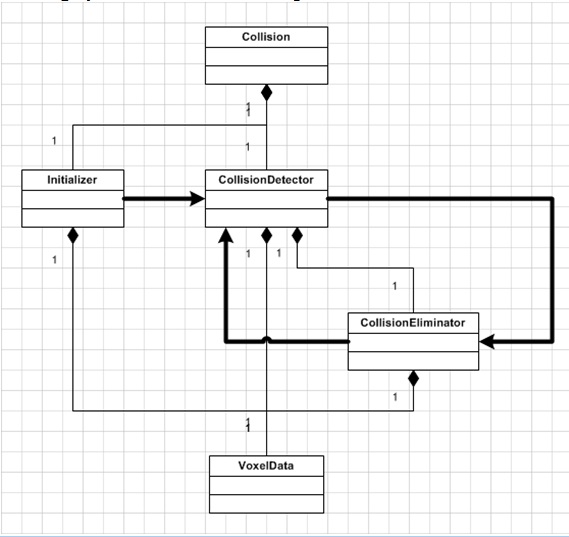}
\caption{\sign{Collision handling system architecture. Black bold arrows shows process direction.}}
\label{overflow}
\end{figure}

\textbf{Collision} class is the wrapper for system. It is mainly responsible for given unified interface.

\textbf{Voxel Data} class is responsible for voxel data and model links for which collision handling is executed.

Class \textbf{Initializer} creates voxel model. Voxel model has been always created for two penetrating objects only. Moreover the model is linked to one of them. During object moving voxel model is also moving relative to object bounding box center. So it is enough to build such voxels only for the contacting half of object.

Interface \textbf{CollisionDetector} describes classes that are responsible for collision detecting. This process of detecting is circumstantially described in the previous chapter.

Interface \textbf{CollisionEliminator} describes classes that eliminate collisions. The method based on additional forces was selected for this task. All methods based on restrictions are improper for this case because collision joint bound calculation is hard and costs much time. Against this we need the quickest collision handling system.

Accordingly to third Newton law the forces acted on collision points has the same magnitude and opposite direction. So we need to define only single direction and one magnitude value. We select direction along straight line that connects both points. The magnitude value was defined experimentally for every task. The only rule for this value is to be larger than object deformation forces.

\section{Results}

\hspace{6mm}A several scenarios were tested after program unit implementation. Both interacting bodies were modeled with nonlinear mass-spring models. Appendix 1 shows these results.

\begin{figure}[ht!]
\centering
\includegraphics[width=90mm]{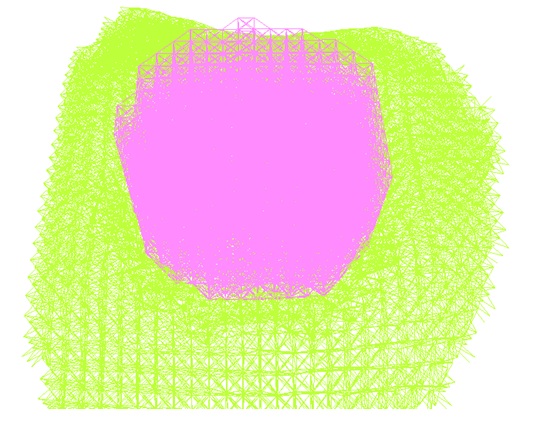}
\caption{\sign{Collision between cylinder and soft body layer.}}
\label{overflow}
\end{figure}

The pictures show that both bodies undergo deformations in collision (Pic. 5, 6 Appendix 1). The algorithm successfully detects and eliminates these collisions. Sometimes separate points are in penetration but this is normal since additional forces can eliminate collision during several iterations.

\begin{figure}[ht!]
\centering
\includegraphics[width=90mm]{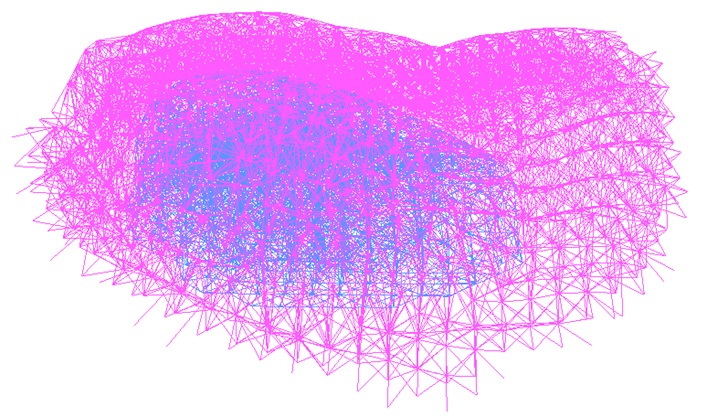}
\caption{\sign{Implant expansion under soft tissue layer.}}
\label{overflow}
\end{figure}

In implant expansion case eliminating with the selected direction does not work very well in practice. Thus to find a direction for given point we take three nearest points from the given and neighbors voxels. Then we build plane on these points and new direction along the plane normal. Also for all three found points we add the third of the force magnitude. The results are presented on Pic. 6 and in Appendix 1.

\section{Conclusion}

\hspace{6mm}This work is about creating program unit that can detect and eliminate collisions in real time. For collision handling it uses modified voxel model. Compared with previous works this model is created only for area of interest. As well as the common voxel model it allows simple collision handling. Moreover this model stores information in voxels about points from one of penetration bodies. This information is used to eliminate collisions simply. If voxels are built for large space area near penetrating bodies then collisions are detected for any deformations.

The program unit was integrated in physical modeling framework to process interaction between implant and soft bodies \cite{nikolaev}.

In future works we are going to improve collision elimination method using information about points in voxels.

\newpage
\section*{Appendix 1. Collision modeling}

\begin{figure}[ht!]
\centering
\includegraphics[width=120mm]{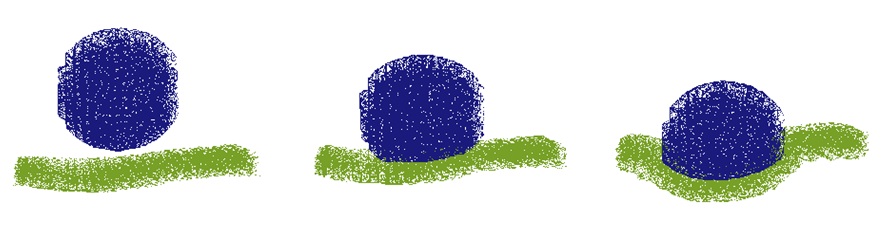}
\caption{\sign{Sphere collision modeling.}}
\label{overflow}
\end{figure}

\begin{figure}[ht!]
\centering
\includegraphics[width=120mm]{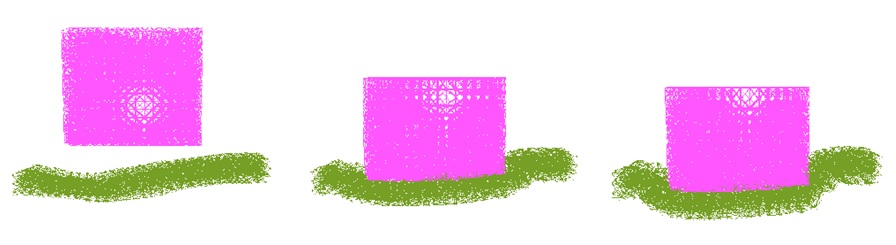}
\caption{\sign{Cylinder collision modeling.}}
\label{overflow}
\end{figure}

\begin{figure}[ht!]
\centering
\includegraphics[width=120mm]{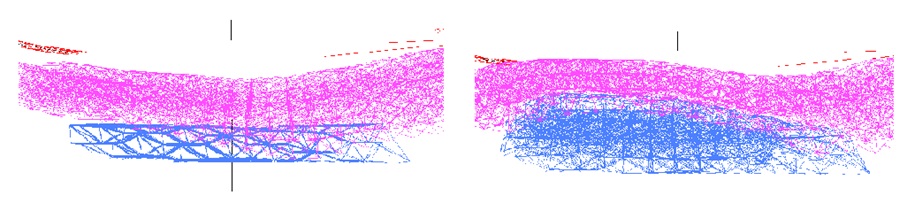}
\caption{\sign{Implant expansion modeling.}}
\label{overflow}
\end{figure}

\end{document}